# Burst statistics of fluctuations in a simple magnetized torus configuration

K. Rypdal, T. Živkovic, L. Østvand
Department of physics and technology, University of Tromsø

In a toroidal plasma confined by a purely toroidal magnetic field the plasma transport is governed by electrostatic turbulence driven by the flute interchange instability on the low-field side of the torus cross section. In this paper we revisit experimental data obtained from the Blaamann torus at the University of Tromsø. On time-scales shorter than the poloidal rotation time, the time series of potential and electron density fluctuations measured on stationary Langmuir probes essentially reflect the spatial poloidal structure of the turbulent field (Taylor hypothesis). On these time scales the signals reveals an intermittent character exposed via analysis of probability density functions and computation of multifractal dimension spectra in different regimes of time scales. This intermittency is associated with the shape and distribution of pronounced spikes in the signal. On time scales much longer than the rotation period there are strong global fluctuations in the plasma potential which are shown to to be the result of low-dimensional chaotic dynamics.

A useful classification of complex systems behavior should allow different behaviors to be distinguished through observation. For instance, in some cases observational spatiotemporal information is only available as fluctuations of one or more point observations, or at best as a scalar two-dimensional (2D) radiation field emitted from a system, while the internal 3D dynamics is not practically accessible to direct observation. Analysis of such fields reveals the existence of complex systems which at the same time exhibit statistical signatures that are traditionally attributed to either turbulence or self-organized critical avalanching dynamics (SOC). Examples are solar flare activity, polar aurora, fluctuations associated with anomalous transport in magnetic fusion plasma, and vortical fluid flows in ordered structures like dusty plasma quasi-crystals. For turbulent systems structure functions of the velocity field reveal an approximate scale-invariant structure, but the deviation from scaling ascribed to intermittency (or multifractality) has received a lot of attention in recent years. For a system in an SOC state local avalanche analysis reveals that the probability distributions of avalanche size and duration are approximately scale-free, but deviations from perfect scaling and their relation to the finite system size have been studied in different frameworks like finite-size scaling and multifractal scaling [1]. Appearance of intermittency has often been taken as a signature of turbulence. For instance, Consolini [2] interpreted multifractality in the time series of the auroral electrojet index as a signature that turbulence plays a major role in the magnetosphere-ionosphere interaction, and recently Govolchanskaya et al. [3] draw the same conclusion from observing spatial intermittency in auroral images. On the other hand, in the fusion literature, mono-fractal scaling with long-range time-dependence has usually been associated with SOC. This calls for a detailed search for multifractal structure in observational data as well as in data from simulations.

In this paper we shall demonstrate that multifractality, monofractality and low-dimensional chaos may appear in the same data set, depending on the scales considered. We revisit data from a previous experiment where time series from the Blaamann device at the University of Tromsø were examined with particular emphasis on detection of long-range temporal dependence in the data [4]. The experimental set-up is described in detail there. Here we only give a brief resumé of the most important points. The plasma is generated by a hot filament discharge in a plasma torus with a purely toroidal magnetic field (no poloidal field component). On the low-field side (LFS) of the torus cross section the electron pressure gradient is directed in the opposite direction of the radius of curvature vector of the magnetic field, and the plasma is unstable to the electrostatic flute interchange instability here. Since hot electrons are injected through the filament in the center of the torus cross section there will be a potential minimum in the center, and this makes the plasma rotate poloidally by $E \times B$ drift. This rotation velocity increases with decreasing neutral gas pressure, and for low neutral pressures the lowest poloidal mode number dominates to give a well defined peak in the power spectrum of signals measured on a stationary probe. The frequency of this peak corresponds to the rotation frequency $f_{rot}$. For high neutral pressure the rotation frequency is smaller than the characteristic time for growth and nonlinear cascading of the flute modes, so this ``turbulent'' regime can be observed for $f \geq f_{rot}$. The spectrum for $f = f_{rot}$ is associated with global fluctuations of the plasma state, and can be considered as nonlinear oscillations in an electric circuit where the plasma plays the role of a strongly nonlinear circuit element.

Fluctuations in plasma potential and electron density measured by electrostatic probes in the middle of the pressure slope on the LFS are shown in Fig. 1. The signals are sampled at 100 kHz, i.e. the sampling time is $\Delta t = 10$ $\mu$s. The plasma rotation period of 2 ms thus corresponds to $200\Delta t$ and is seen as a weak bump at 0.5 kHz on the potential power spectrum in Fig. 2, and as a stronger bump on the electron density spectrum. The most prominent feature observed by direct examination of the signals in Fig. 1 is the prominent spikes occuring with interspike intervals ranging from $20-200$ $\Delta t$. This time range corresponds to the plateaus observed in the frequency range $0.5-5$ kHz in Fig. 2. The spikes are very peaked, with width $< 20\Delta t$. These time scales correspond to the power-law tail $S(f): f^\beta$ of the spectra for $f > 5$ kHz, with $\beta \approx 2.5$. On these short time-scales the signals are smooth, so the fractal dimension of their graph at these scales is $D = 1$. If the process were self-affine this would correspond to a Hurst exponent $H = 2 - D = 1$, and a spectral exponent $\beta = 2H + 1 = 3$. However, as we shall see below, the process is far from self-affine, but it is easily shown that the power spectrum of randomly distributed spikes with exponential rise and decay will have a Lorenzian power spectrum, which has a power-law tail with $\beta = 2$. Thus the sharp peaks of the spikes are responsible for this part of the power spectrum and, as will be seen below, for the multifractal (or intermittent) nature of the signals on this short time scale. The flat plateaus in the spectra in the frequency range $0.5-5$ kHz reflect the white-noise character of the distribution of the spikes, and are similar for potential and density. In fact, the larger potential and density peaks are correlated, positive density spikes are associated with negative potential spikes, as they should for flute modes. For frequencies below 1 kHz the electron density spectrum (apart from the rotation hump) has a more or less flat white noise like spectrum, whereas the potential has a power-law range for $0.1 < f < 1$ kHz with $\beta \approx 1.4$, which is indicative of a strongly antipersistent self-affine motion with $H = (\beta-1)/2 \approx 0.2$. Since these low-frequency fluctuations are not present in the electron density spectrum, and are on frequencies below the plasma rotation frequency, they are not associated with flute-mode turbulence, but rather to global oscillations, as mentioned above. For frequencies below 0.1 kHz this spectrum becomes flat, indicating a white-noise like process with complete loss of temporal correlation on time scales longer than 10 ms ($1000\Delta t$).

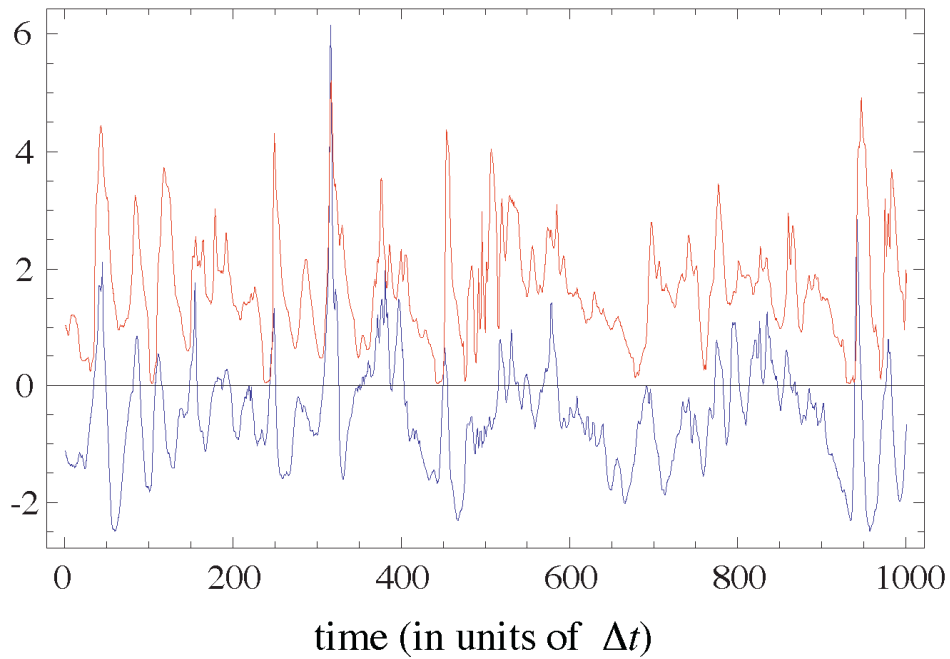

Figure 1: Upper trace (red) is the time series of electron density fluctuations. Lower trace (blue) is the same for plasma potential. Units are arbitrary, and the plasma potential has an arbitrary zero potential.

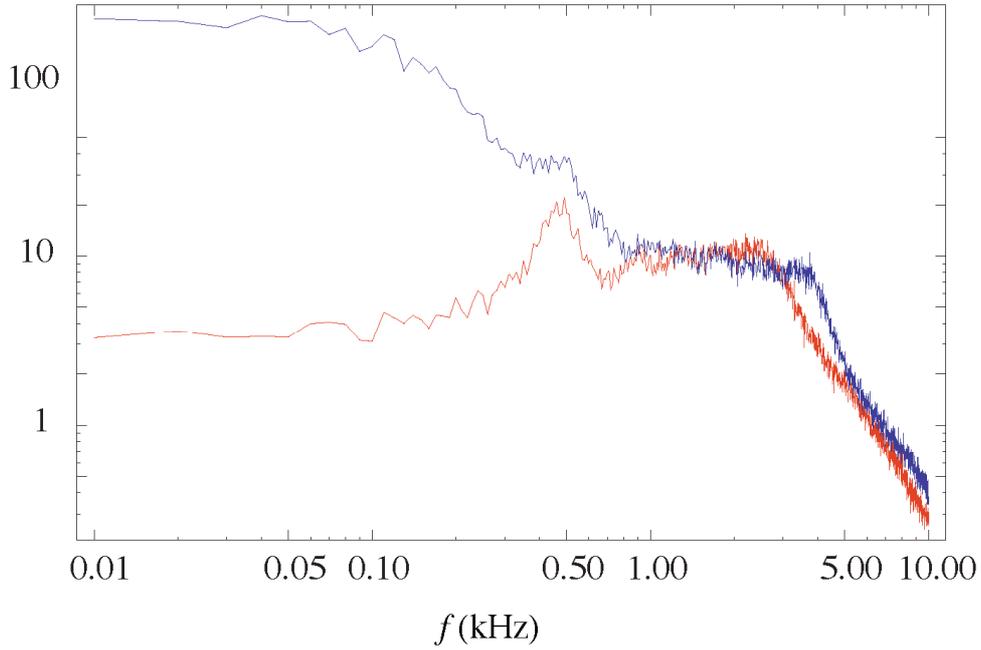

Figure 2: Log-log power spectra for electron density (lower red curve) and for plasma potential (upper blue curve).

The probability density functions (PDFs) of differences in potential and density separated by a time interval $\tau$ are shown for varying $\tau$ in Fig. $3a$ and $3c$. Both distributions are leptokurtic for small $\tau$, but while the electron density rapidly converges towards a near-gaussian shape (which has kurtosis $K=3$), the plasma potential converges to a leptokurtic distribution with $K \approx 10$, as shown in Fig. 4. The increasing kurtosis at small scales is a fundamental statistical signature of intermittency.

On the long time scales, where the signals have a stationary character, it makes more sense to consider the PDFs of the coarse grained signals [4]. Such signals are made by running averages over intervals of length $\tau$. Such PDFs for increasing degrees of coarse-graining are shown in Fig. $3b$ and $3d$, and show a near-gaussian distributions for density, while the potential maintains a leptokurtic and strongly skewed distribution throughout the scales corresponding to frequencies $f > 0.1$ kHz. These PDFs maintain relatively constant shape through the scales corresponding to the power-law frequency range $0.1 < f < 1$ kHz, and hence the signals have a self-affine (and hence monofractal) character in this range. In the range $f < 0.1$ kHz the signals are more gaussian-like, and has the character of gaussian white noise.

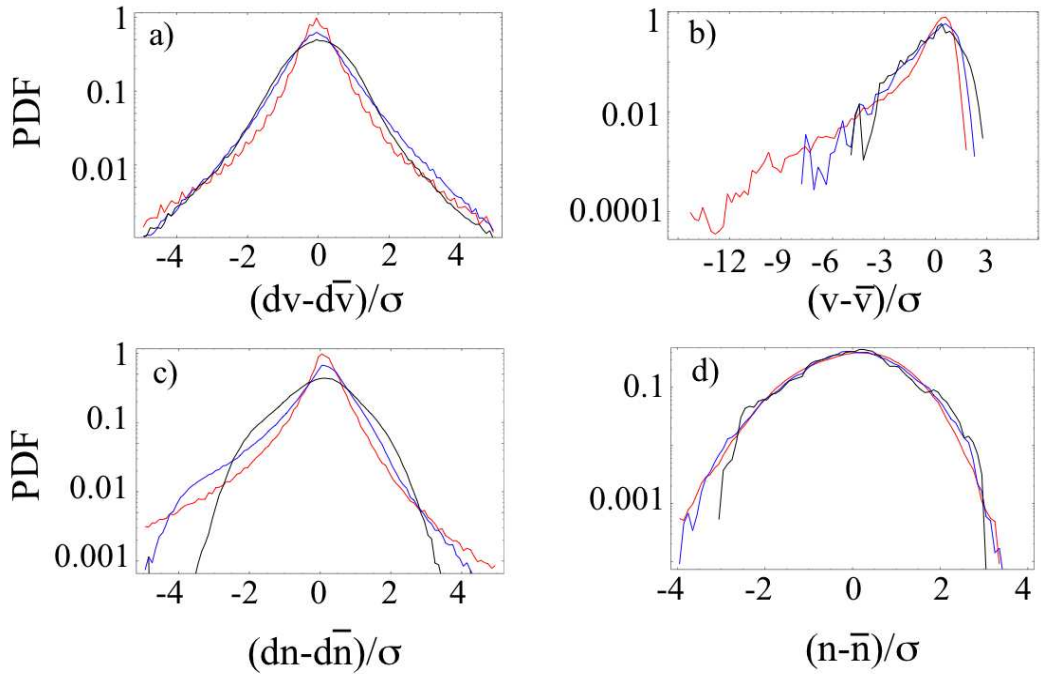

Figure 3: Logarithmic plots of PDF for a) differences in plasma potential separated by time $\tau$, where red is $\tau=2^0$, blue is $\tau=2^3$ and black is $\tau=2^6$; b) coarse-grained plasma potential averaged over interval $\tau$, where red is $\tau=2^8$, blue is $\tau=2^{11}$ and black is $\tau=2^{14}$; c) differences in the plasma density, where red is $\tau=2^0$, blue is $\tau=2^3$ and black is $\tau=2^6$ (larger scales are the same as $\tau=2^6$); d) coarse-grained electron density, where red is $\tau=2^8$, blue is $\tau=2^{11}$ and black is $\tau=2^{14}$. Time $\tau$ is in units of $\Delta t$.

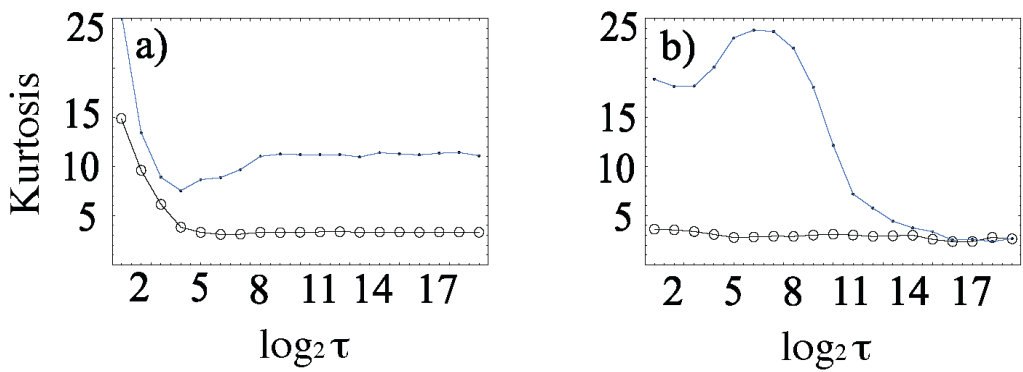

Figure 4: a) Kurtosis for differences in plasma potential separated by time $\tau$ (dotted line) and for differences in electron density (open circles). b) Kurtosis for coarse-grained plasma potential averaged over interval $\tau$ (dotted line), and for coarse-grained electron density (open circles).

The rapid fall of kurtosis in Fig. 4 occurs over time scales $\tau < 2^5$, corresponding to frequencies in the power-law tail of the spectrum. This is the time scale of the peaks of the spikes, and thus the drop of the kurtosis (intermittency) is a property that characterizes the shape and distribution of these spikes. A systematic characterization of this intermittency can be done by computation of the multifractal spectrum of dimensions $D_q$ (see e.g. [2]). This can be done when one differenciates the time-series, then takes the absolute value of each increment and considers this as a ``mass density distribution'' in time. We consider how the integrated ``mass'' in boxes of length $\tau$ is distributed for varying $\tau$ (again a coarse-graining procedure) and compute statistical moments of increasing order $q$ for each $\tau$. For a self-affine (monofractal) signal these moments scale as : $\tau^{(q-1)D_q}$, where $D_q = D_B$, where $D_B$ is the box fractal dimension of the mass distribution. In this case there is some ``mass'' in every time point (there are no empty ``holes'' in the mass distribution), so $D_B = 1$. For a multifractal signal $D_q$ is a monotonically decreasing function which varies between two limits $D_{-\infty} > D_{\infty}$, where $D_{-\infty} - D_{\infty}$ measures the degree of intermittency. Specifically $D_0$ is the box dimension $D_B$, $D_1$ is known as the information dimension, and $D_2$ as the correlation dimension. This formalism, however, is based on the assumption that the moments exhibit power-law scaling as a function of $\tau$. In our data this scaling is not perfect, but we can identify three different power-law regimes corresponding roughly to the scales $\tau : (1-16)\Delta t$, $\tau : (16-128)\Delta t$, and $\tau > 128\Delta t$. This means that we can define multifractal spectra $D_q$ for each regime of scales. These spectra are shown in Fig. 5, and demonstrate that multifractality and intermittency are properties associated with the small scales that resolve the structure and distribution of the narrow spikes in the signal.

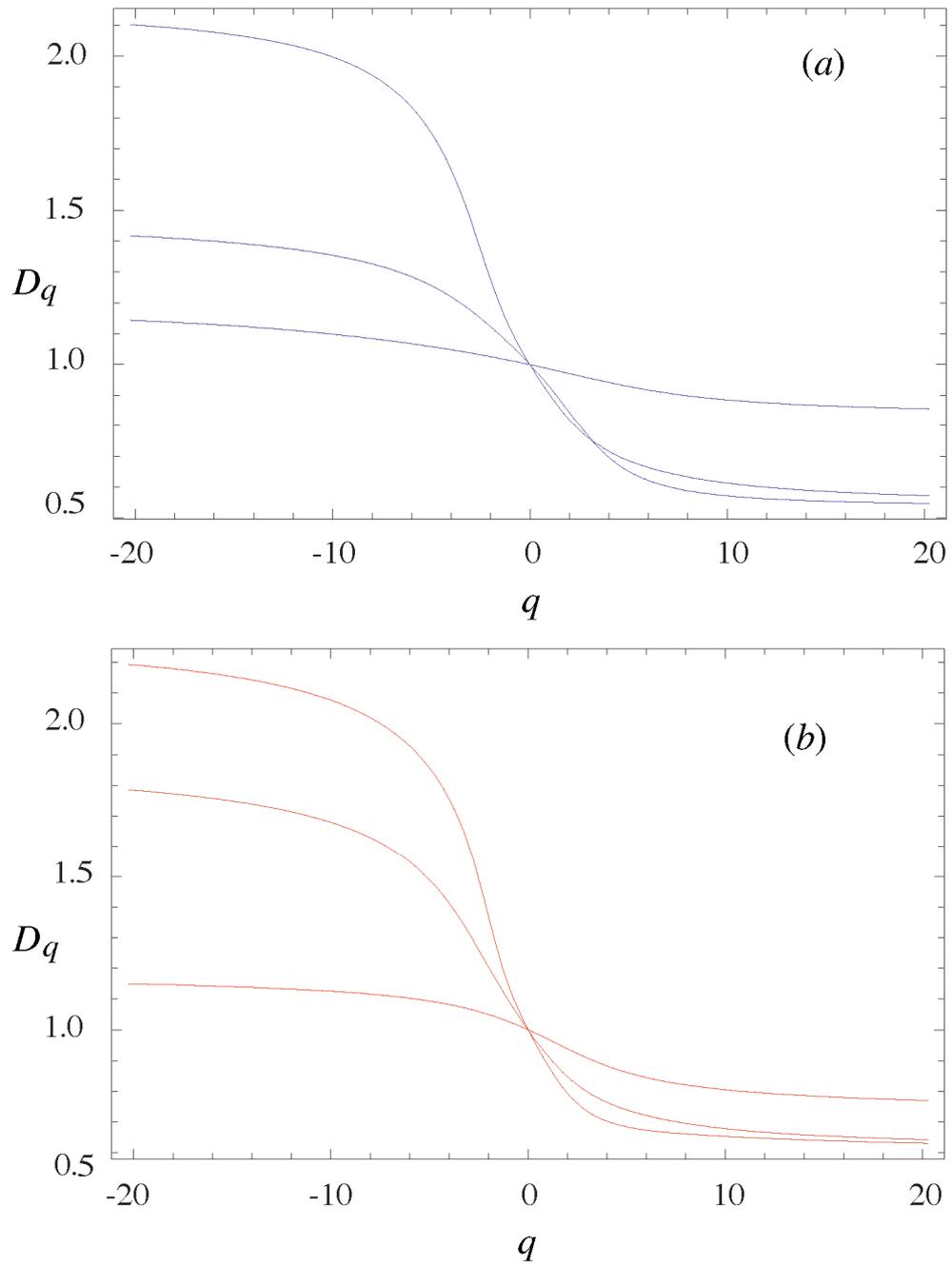

Figure 5: Multifractal dimension spectra for a) plasma potential and b) electron density. The spectra are obtained from fitting straight lines to the log-log plot of the moment functions in certain intervals of the time separation $\tau$. For the upper curves in both panels the interval $1 < \tau < 16$ was used, for the middle curves $16 < \tau < 128$, and the lower curves $128 < \tau < 4096$.

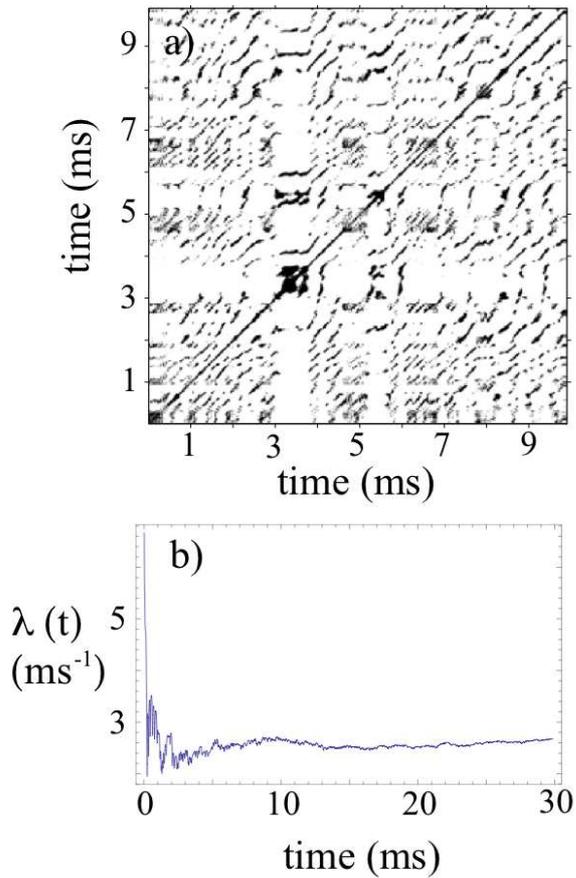

Figure 6: Chaotic characteristics of the filtered plasma potential: a) Recurrence plot, b) Lyapunov exponent.

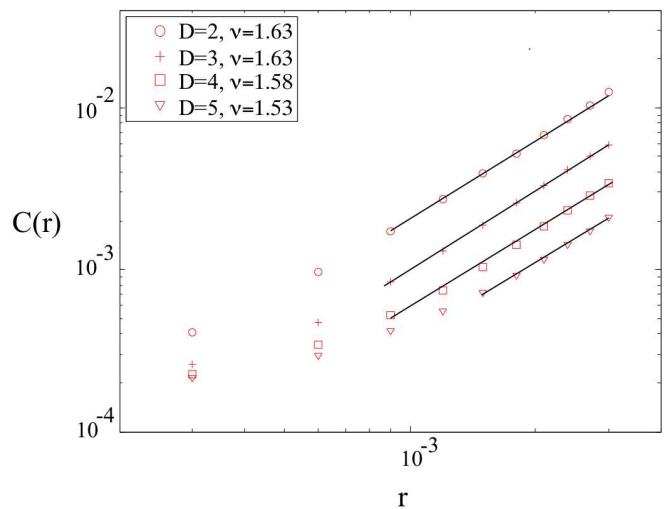

Figure 7: Log-log plot of correlation integrals $C(r)$ for embedding dimension $D = 1,2,3,4,5$. The full lines are linear fits to the curves, which all have slopes $\nu \approx 1.6$.

So far, what we know about the low-frequency fluctuations is that they are uncorrelated and near-gaussian. It would be interesting to know if they should be considered as stochastic, or if they may be a result of low-dimensional chaotic dynamics. For this purpose we apply a mexican hat wavelet filter on the plasma potential in order to extract the dynamics on scales larger than the poloidal rotation time, and hence wavelet coefficients corresponding to frequencies larger than the rotational frequency are set to zero by the filter. For this low-pass filtered signal we compute the largest Lyapunov exponent $\lambda$ for the plasma potential, which should be positive if the dynamics is chaotic. In Fig. 6$b$, the coefficient of exponential divergence of neighboring phase-space trajectories are computed for increasing time, and converges to the largest Lyapunov exponent $\lambda(t): 2 \times 10^{-3}$ s$^{-1}$. However, it has been shown in [5] that stochastic time series can also give positive Lyapunov exponents when calculated numerically. Therefore, we apply recurrence plot analysis (see [6]), which visualizes the phase space dynamics.

Prior to constructing a recurrence plot the phase space is reconstructed by time-delay embedding [7], where vectors $x_i$ ($i = 1,...,T$) are produced. Then a $T \times T$ matrix consisting of elements 0 and 1 is constructed. The matrix element $(i, j)$ is 1 if the distance is $|x_i - x_j| \leq r$ in the reconstructed phase space, and otherwise it is 0. The recurrence plot is simply a plot where the points $(i, j)$ for which the corresponding matrix element is 1 is marked by a dot. The radius $r$ is fixed and chosen such that a sufficient number of points are found to reveal the fine structure of the plot. Short diagonal lines indicate a chaotic state where trajectories recur for a short time and then diverge exponentially again. In Fig. 6$a$, we see that the low-pass filtered plasma potential fluctuations exibit chaotic signatures.

In order to determine the correlation dimension of the chaotic attractor, we first calculate the modified correlation integral

$$C(r,W) = \frac{2}{N^2} \sum_{n=W}^{N} \sum_{i=1}^{N-n} H(r - |x_{i+n} - x_i|)$$

introduced in [8]. Here $H(...)$ is a Heaviside function, $r$ is the radius of a ball in phase space, and $W$ is approximately equal to the autocorrelation time of the time series. Further, we calculate the correlation dimension $\nu$ from $C(r,W) = r^\nu$ for various $r$. For the low-pass filtered potential, we plot $C(r,W)$ vs. $r$ for the embedding dimensions $D = 2, 3, 4$ and $5$ in Fig. 7. As one can see, the correlation dimension changes very little for different $D$, which further implies that chaotic attractor for the low-pass filtered potential has dimension $\nu \approx 1.6$, and hence that the chaotic dynamics is low-dimensional.